\documentclass{edm_template}
\usepackage{float}
\begin{document}

\title{Identifying Student Communities in Blended Courses}
%
%
%
%
%

\numberofauthors{1} 
%
\author{
%
%
\alignauthor
Niki Gitinabard, Collin F. Lynch, Sarah Heckman, Tiffany Barnes\\
       \affaddr{North Carolina State University}\\
       \affaddr{Computer Science Department}\\
       \affaddr{Raleigh, NC, US}\\
       \email{\{ngitina, cflycnh, sarah\_heckman, tmbarnes\}@ncsu.edu}
}


\maketitle
\begin{abstract}
Blended courses have become the norm in post-secondary education.  Universities use large-scale learning management systems to manage class content.  Instructors deliver readings, lectures, and office hours online; students use intelligent tutors, web forums, and online submission systems; and classes communicate via web forums.  These online tools allow students to form new social networks or bring social relationships online.  They also allow us to collect data on students' social relationships. In this paper we report on our research on community formation in blended courses based on online forum interactions. We found that it was possible to group students into communities using standard community detection algorithms via their posts and reply structure and that the students' grades are significantly correlated with their closest peers. 

\end{abstract}

%


\section{Introduction}
Improvements in technology have facilitated new models of student and instructor engagement. Students now supplement the traditional course structure with online materials.  Instructors can share class material online, have an online discussion forum, or make quizzes and homework submissions online.  This in turn provides a wealth of new data on student behaviors that we can use to study students' social relationships. In particular it allows us to study the impact of these social ties on course outcomes. 

In prior work Brown et. al. showed that students in MOOCs form pedagogically-relevant, and homogeneous social networks. Brown et. al. has shown that students can be clustered into stable communities based upon their pattern of online questions and replies \cite{brown15w}. They have also shown that students' final grades are significantly correlated with those of their closest peers and community group.  They have also shown that these communities, while homogeneous in terms of performance, are not united by their incoming motivations for enrolling in the course nor for their prior experience level \cite{brown15}. 

To date these results have only been found in MOOCs where the user forum represents students' primary connection to one-another, and almost all relevant course interactions occur online. Students in blended courses, by contrast, often have preexisting social ties that carry over from prior courses at the same institution. In this paper we show that while forum interactions are not the only means of communication between students, they still define the same communities as was found in MOOCs and that the students' final grades are significantly correlated with those of their community members.

\section{Dataset Information}

In this paper we report on studies of three distinct courses, ``Discrete Math-2013'', ``Discrete Math-2015'' and ``Java Programming Concepts-2015''. All three are undergraduate computer science courses, offered  at NC State and include significant blended components. Discrete Math-2015 and Java Programming Concepts-2015 occurred contemporaneously during the Fall 2015 semester while Discrete Math-2013, a previous offering of Discrete Math-2015, was offered in Fall 2013.

\section{Methods}

\subsection{Defining Social Interactions}
Each node in our social networks represents an individual participant in the class. In the first class anonymous posting was allowed, so we have an unknown user related to all the anonymous posts.  Social relationships are represented as arcs. We define a social relationship based upon direct and indirect replies in the user forum.  Our method was similar to that of Brown et. al. \cite{brown15}. We defined an edge between A and B if B replied to a thread after A had done so. This interaction can include starting the original thread, replying with a follow-up, or posting a feedback on a reply. We then aggregate these edges to form a weighted graph containing arcs for all of the relations. We assume that anyone who posts on a thread has read the prior comments before doing so.  Thus it defines a form of social interaction between the participants as the students are expressly choosing to make a public reply to one another. For the purposes of the present analysis we included only students in our network and thus confined our social relationships to between-student connections.

\subsection{Graph Analysis}
For each of the graphs we generated, we removed the isolated vertices and performed clustering using the method described in \cite{brown15, brown15w}. Our clustering method is an iterative process where we evaluate the modularity of graphs with an increasing number of clusters until we find a limit point where the modularity almost stops growing, which indicates the \textit{natural cluster number}. After finding the natural number, on each iteration we generated the clusters via the Girvan-Newman edge-centrality algorithm\cite{girvan02}.  On each iteration the algorithm removes the most central edge and and repeats until a set of k disjoint clusters has been produced. We then assessed whether or not the grade distributions in different clusters are significantly different by calculating the Kruskal-Wallis (KW) correlation between cluster assignment and grade.  Kruskal-Wallis is a nonparametric analogue to the more common ANOVA test \cite{kruskal52}.

\section{Results}

\begin{figure}
  \centering

  \includegraphics[width=0.4\textwidth]{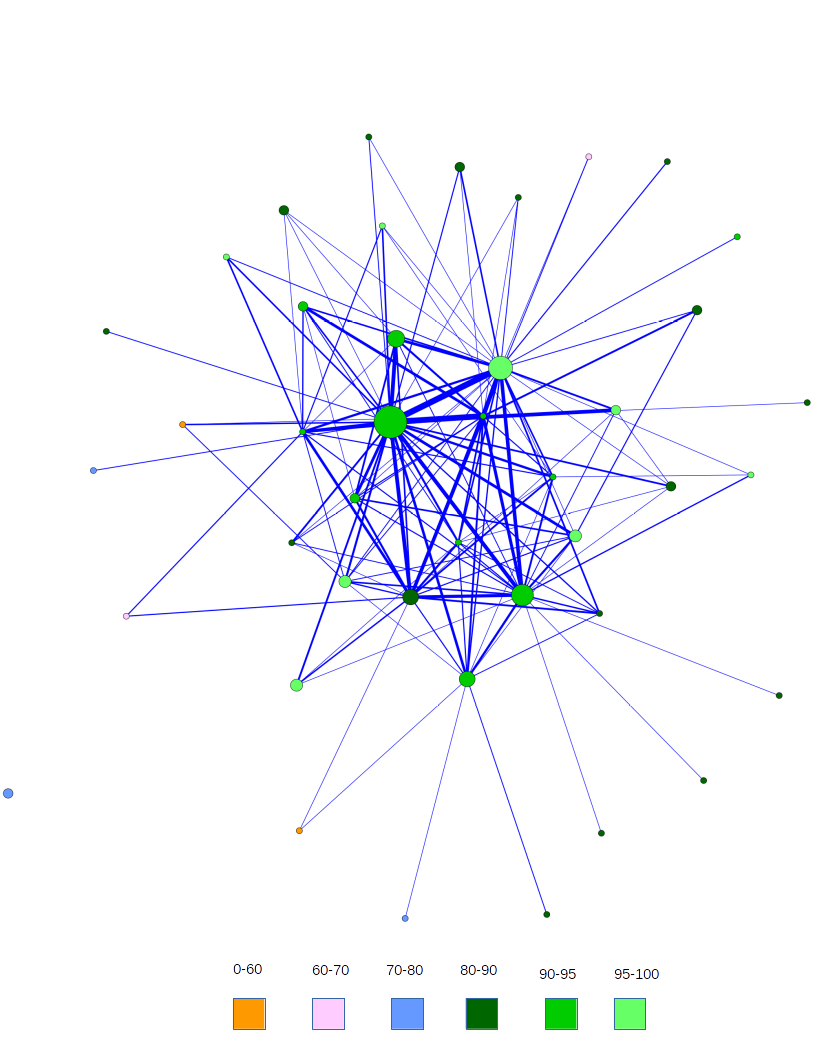}
   \caption{Communities generated on Discrete Math 2013 class
}
    \label{barnes14_piazza_only}
\end{figure}

In graphs generated for Discrete Math 2014, we found that the graph reaches its natural cluster number at 42. We performed the Girvan Newman clustering and the resulting clusters can be seen in Figure~\ref{barnes14_piazza_only}. In this graph, each node represents a community, the size of the nodes shows the number of members and the color shows their average grade. We can observe that the KW correlation between cluster number and the grades is statistically significant ( p = 0.044 < 0.05 ), which is similar to the results in MOOCs.

Our results show that, for Discrete Math 2015 ( p = 0.004 < 0.05 ) and Java Programming Concepts 2015 ( p = 0.015 < 0.05 ) graphs, there is a similar significant KW correlation between student grades and their communities.



\section{Discussion, Conclusions and Future Work}

In this paper, we generated a social graph between students in three different blended courses based on forum interactions. We found that similar to MOOCs, communities are formed in these graphs whose members tend to have similar grades. This is consistent with prior work which indicates that student communities on forum may be used to predict course outcomes \cite{brown15w, brown15}. 

Having access to these social graphs can help instructors to identify the communities formed among students which can be used to  find the students who need more help earlier. Our research does not show causality. Thus more research is needed to find out whether being in the communities makes their grades similar, or students are just likely to interact with others who are more like them. If we find out that the community membership has an effect on students' performance, we can use this information to identify isolated or poorly-performing groups early in the course and intervene by encouraging them to make contact with better students or seek help as a group.

There has been much work done on how forum interactions in MOOCs, being a hub in a social network or how being at the center of the graph could affect students' performance. We can use these graphs to conduct more research on which interaction levels will lead to better grades.

In further work we plan to address whether or not we can identify other types of social ties in blended courses, since the communications are more complicated.


\section{Acknowledgments}
This work was supported by NSF grant \#1418269: ``Modeling Social
Interaction \& Performance in STEM Learning'' Yoav Bergner, Ryan
Baker, Danielle S. McNamara, \& Tiffany Barnes Co-PIs.

%
\bibliographystyle{abbrv}
\bibliography{references}
%
%

\end{document}